\documentclass[preprint,aps,amsmath,byrevtex,prd,titlepage,
nofootinbib,12pt]{revtex4-2}

\usepackage{dcolumn}
\usepackage{amsmath}
\usepackage{amssymb}
\usepackage{bm}
\usepackage{hyperref}
\usepackage{graphicx}
\usepackage{slashed}

\newcommand{\beq}{\begin{equation}}
\newcommand{\eeq}{\end{equation}}
\newcommand{\eq}[1]{Eq.~(\ref{#1})}

\begin{document}

\title{One-Loop Electron Mass and QED Trace Anomaly}

\author {Michael I. Eides}
\email[Email address:] {meides@g.uky.edu}
\affiliation{Department of Physics and Astronomy,
University of Kentucky, Lexington, KY 40506, USA}
%\date{}

\begin{abstract}
Electron mass is considered as a matrix element of the energy-momentum trace in the rest frame. The one-loop diagrams for this matrix element are different from the textbook diagrams for the electron mass renormalization. We clarify connection between the two sets of diagrams and explain analytically and diagrammatically why the results of both calculations coincide.
\end{abstract}

\maketitle

%\tableofcontents

\section{Introduction}

Hadron energy-momentum tensor (EMT), its matrix  elements, anomalous trace, form factors and multipole expansion is now a vibrant field of research. EMT form factors describe interaction of particles with weak external gravitational field \cite{Kobzarev:1962wt,Pagels:1966zza}. For a long time there was no way to measure form factors of hadron EMT, but the situation changed when it was realized that they are connected to the generalized parton distribution functions which can be measured in deeply virtual Compton scattering and other hard exclusive reactions, see, e.g.,  \cite{Ji:1996ek,Ji:1996nm,Radyushkin:1996ru,Collins:1996fb,Kharzeev:1998bz,Berger:2001xd}.

Due to nonperturbative nature of the low-energy QCD theoretical studies of hadron gravitational form factors and other hadron EMT properties use general gauge theory principles, lattice QCD, QCD inspired low-energy models and model theories which admit quantitative analysis, see e.g., \cite{Ji:1994av,Ji:1995sv,Hudson:2017xug,Polyakov:2018zvc,Kharzeev:2021qkd,Liu:2021gco,Lorce:2021xku,Ji:2020bby} and numerous other papers.

A new insight into the EMT properties could arise from consideration of  EMT in theories which allow  perturbative treatment. While perturbative approach is clearly impossible in QCD, one can consider a simpler gauge theory, namely QED, and hope to acquire some experience which would be useful for the hadronic world. One-loop QED contributions to the EMT
form factors, matrix elements and trace were calculated for a free electron and electron in the Coulomb field in a number of old and recent papers \cite{Milton:1971xnd,Milton:1973zz,Berends:1975ah,Milton:1976jr,Ji:1998bf,
Rodini:2020pis,Sun:2020ksc,Ji:2021qgo,Metz:2021lqv,Ji:2021mfb,Ji:2022exr,Freese:2022jlu}.

One-loop electron mass renormalization in the mass-shell renormalization scheme is a textbook problem discussed in every introductory quantum field theory textbook, see, e.g., \cite{Peskin:1995ev}. Consideration of EMT suggests another perspective on this classical problem. One can calculate electron mass as a matrix element of the EMT trace. The diagrams describing this matrix element do not coincide with the well known diagrams for the one-loop corrections to the electron mass. We will calculate electron mass with the help of both sets of different contributions and explain why they produce coinciding results.

\section{Matrix Elements of EMT and Mass of Particles}

General formulae for EMT $T^{\mu\nu}(x)$ follow from its definition as a conserved two-index symmetric tensor. Due to translational invariance

\beq \label{lorinvmael}
\langle\bm p'|T^{\mu\nu}(x)|\bm p\rangle=e^{i(p'-p)\cdot x}
\langle\bm p'|T^{\mu\nu}(0)|\bm p\rangle,
\eeq

\noindent
where $|\bm p\rangle$ is a particle eigenstate with momentum $\bm p$ and states here are normalized relativistically, $\langle\bm p'|\bm p\rangle=2E_p(2\pi)^3\delta^{(3)}(\bm p-\bm p')$.

The Hamiltonian  is a three-dimensional integral, $H=\int d^3xT^{00}(x)$, and on the one hand

\beq \label{hamamtrel}
\int d^3x\langle\bm p'|T^{00}(x)|\bm p\rangle=2E_p^2(2\pi)^3\delta^{(3)}(\bm p-\bm p'),
\eeq

\noindent
and on the other hand (see  \eq{lorinvmael})

\beq
\int d^3x\langle\bm p'|T^{00}(x)|\bm p\rangle=(2\pi)^3\delta^{(3)}(\bm p-\bm p')\langle\bm p'|T^{00}(0)|\bm p\rangle.
\eeq

\noindent
Hence,
\beq
\langle\bm p|T^{00}(0)|\bm p\rangle=2E_p^2,
\eeq

\noindent
and due to Lorentz invariance

\beq \label{lorinvmas}
\langle\bm p|T^{\mu\nu}(0)|\bm p\rangle=2p^\mu p^\nu,\qquad \langle\bm p|{T^\mu}_\mu(0)|\bm p\rangle=2m^2.
\eeq

\noindent
In the rest frame and with the nonrelativistic normalization of states $\langle\bm p'|\bm p\rangle=(2\pi)^3\delta^{(3)}(\bm p-\bm p')$ 

\beq \label{masstrmy}
\langle\bm 0|{T^\mu}_\mu(0)|\bm 0\rangle=m,\qquad  \langle\bm 0|T^{00}(0)|\bm 0\rangle=m.
\eeq

\noindent
These relations hold both elementary particles and for bound states, and are obviously valid in any relativistic  field theory. Below we will consider the first of these equations for an electron in QED.

Symmetric EMT tensor is conserved in a translationally invariant relativistic field theory and it is not renormalized as any conserved operator $T_0^{\mu\nu}=[T^{\mu\nu}]_R$. It is well known that EMT trace in gauge theories acquires an anomalous contribution \cite{Nielsen:1977sy,Adler:1976zt,Collins:1976yq} and has the form

\beq  \label{anomtrac2}
{T_0^\mu}_\mu=(1+\gamma_m(e_0))\bar\psi_0 m_0\psi_0+\frac{\beta(e_0)}{2e_0}F^2_0
=(1+\gamma_m(e))[\bar\psi m\psi]_R+\frac{\beta(e)}{2e}[F^2]_R,
\eeq

\noindent
where $\beta(e)/2e=\alpha/6\pi$, $\gamma_m(e)=3\alpha/2\pi$%, and we attached subscript {\scriptsize $0$} to bare quantities.
.

The left hand side in \eq{anomtrac2} is renorminvariant and then the sum of the operators on the right hand side (RHS) is also renorminvariant. There are subtleties with separation of the terms on the right hand side in a sum of renorminvariant   operators beyond one loop, see  \cite{Tarrach:1981bi,Hatta:2018sqd,Tanaka:2018nae,Metz:2020vxd,Lorce:2021xku,Ahmed:2022adh,Tanaka:2022wzy}.

We are going to consider matrix element  of the anomalous trace in \eq{anomtrac2} for an electron at rest in the one-loop approximation. We will be working in the renormalized perturbation theory and use the  mass-shell renormalization scheme. Then, according to \eq{masstrmy}, this matrix element  should be equal to the physical electron mass $m$ and describe one-loop mass renormalization. At the same time the diagrams which contribute to this matrix element do not coincide with the well known mass renormalization diagrams. Our goal is to clarify from the diagrammatic and analytic perspectives, why two different diagrammatic descriptions lead to the identical results\footnote{One-loop matrix element of the EMT trace in the electron state was calculated in the mass-shell renormalization scheme with the momentum cutoff in \cite{Sun:2020ksc}, but the relationship between two sets of diagrams was not addressed there.}.

\section{One-loop mass renormalization and EMT anomalous trace for a free electron}

Let us recall one-loop electron mass renormalization in the mass-shell scheme with dimensional regularization. We collected the well known relevant formulae in the Appendix. In the mass-shell renormalization scheme the counterterm $\delta m^{(2)}$ kills the ultraviolet divergence in the regularized but not renormalized self-energy diagram $\Sigma(p)$  and preserves the physical mass at $m$

\begin{figure}[h!]
\begin{center}
\includegraphics
[height=1.5cm]
{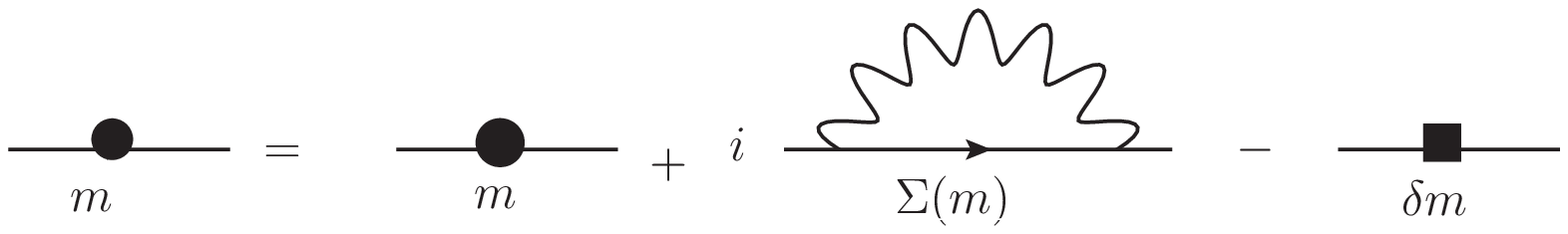}
\end{center}
\caption{One-loop mass renormalization.}
\label{mass_ren_stn}
\end{figure}

\beq \label{massnonren}
m= m+\Sigma(m)-\delta m^{(2)},
\eeq

\noindent
where $\delta m^{(2)}=\Sigma(m)$. This expression is illustrated in Fig.~\ref{mass_ren_stn}. The factor $i$ before the self-energy diagram in Fig.~\ref{mass_ren_stn} is included in the standard diagrammatic definition of $\Sigma(p)$, see e.g., \cite{Peskin:1995ev}.

Next we turn to the matrix element of the EMT trace in \eq{masstrmy} for the electron. The leading contribution to the matrix element of the term $(\beta(e_0)/2e_0)F_0^2$ in \eq{anomtrac2} is of order $\alpha^2$ and we ignore it. It is sufficient to calculate  the one-loop matrix element

\beq \label{matrelonshdinr}
T=\langle\bm  0|m_0(1+\gamma_m)\bar\psi_0\psi_0|\bm 0\rangle.
\eeq

\begin{figure}[h!]
\begin{center}
\includegraphics
[height=1.5cm]
{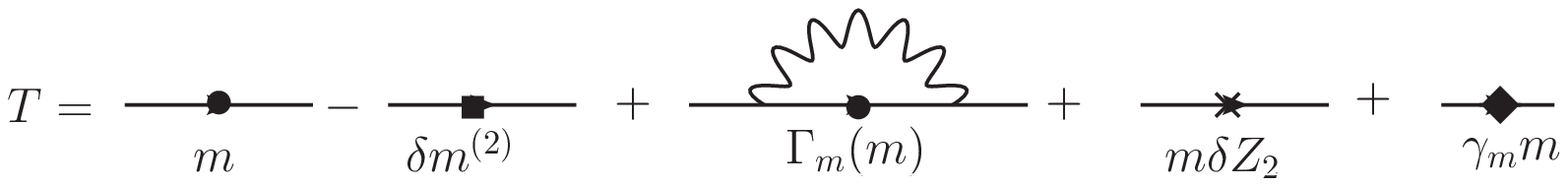}
\end{center}
\caption{One-loop matrix element of the EMT trace.}
\label{tworelsecc}
\end{figure}

\noindent
In the one-loop approximation

\beq \label{ferrelsum}
T=\langle\bm 0|(1+\gamma_m)m_0\bar\psi_0\psi_0 |\bm 0\rangle\approx \langle\bm 0|(1+\gamma_m)(m-\delta m^{(2)})\bar\psi_0\psi_0 |\bm 0\rangle
\eeq
\[
=m+m\delta Z_2-\delta m^{(2)}+m\gamma_m+\Gamma_m(m)
\]

\noindent
where $\Gamma_m(m)$ is the one-loop diagram for the scalar vertex $m\bar\psi\psi$, see Fig.~\ref{tworelsecc}.

\noindent
All terms on the RHS in \eq{ferrelsum} and in Fig.~\ref{tworelsecc}, except $\Gamma_m(m)$, are known from the one-loop mass-shell renormalization scheme (see \eq{mdz2dm} and \eq{deltaz2}) and only $\Gamma_m(m)$ requires calculation. After an easy one-loop calculation we obtain

\beq
\Gamma_m(m)=\frac{\alpha}{4 \pi } m \left\{\frac{8}{\epsilon}-4 \gamma+2 \ln \frac{\lambda^2}{m^2}+4 \ln\frac{\mu ^2}{m^2}+2+4 \ln (4 \pi )\right\},
\eeq

\noindent
where $\lambda$ is the infrared photon mass and $\gamma$ is the Euler constant.

Next we use $\Gamma_m(m)$, the renormalization constants in \eq{mdz2dm} and \eq{deltaz2},  and $\gamma_m=3\alpha/2\pi$ to calculate the sum in \eq{ferrelsum}

\beq \label{sumofford}
T=m+m\delta Z_2-\delta m^{(2)}+m\gamma_m+\Gamma_m(m)=m.
\eeq

\noindent
Thus we confirmed that the matrix element $T$ of the anomalous EMT trace in the one-loop approximation is equal the physical electron mass, as it should be. Comparing \eq{massnonren} and \eq{sumofford} (and the respective Figs.~\ref{mass_ren_stn} and \ref{tworelsecc}) we see that

\beq \label{sigmagamrel}
\Sigma(m)=\Gamma_m(m)+m\delta Z_2+\gamma_mm.
\eeq

\noindent
At this stage it is unclear why the different sets of diagrams in Fig.~\ref{mass_ren_stn} and Fig.~\ref{tworelsecc} produce coinciding results. To figure out a deeper reason why this happens we expand  the unrenormalized electron self-energy $\Sigma(\slashed p)$  in the Taylor series near the physical mass

\beq
\Sigma(\slashed p)=\Sigma(\slashed p=m)+(\slashed p-m)\Sigma'(\slashed p=m)+O((\slashed p-m)^2)
\eeq
\[
=\delta m^{(2)}+(\slashed p-m)\delta Z_2+O((\slashed p-m)^2).
\]

\noindent
Differentiating with respect to $m$ we obtain at $\slashed p=m$

\beq \label{selfenexpdif}
m\frac{d\Sigma(\slashed p)}{dm}_{|\slashed p=m}=m\frac{d\Sigma(\slashed p=m)}{dm} -m\Sigma'(\slashed p=m).
\eeq

\begin{figure}[h!]
\begin{center}
\includegraphics
[height=1.5cm]
{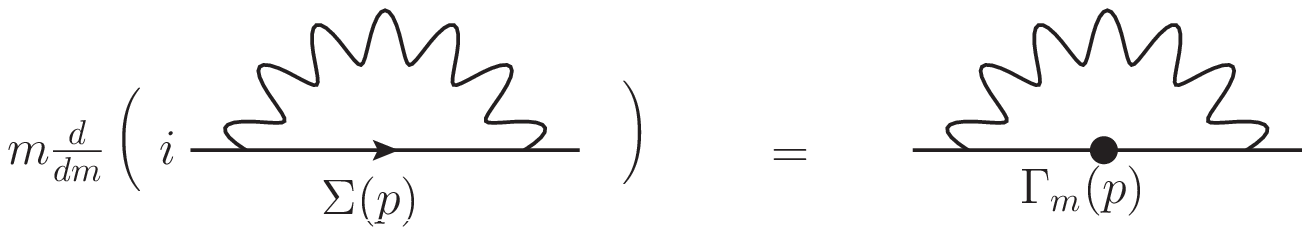}
\end{center}
\caption{Logarithmic mass derivative of $\Sigma(p)$.}
\label{deriv}
\end{figure}

\noindent
Notice that (see Fig.~\ref{deriv})

\beq \label{gammdersigm}
m\frac{d\Sigma(p)}{dm}=\Gamma_m(p),
\eeq

\noindent
which holds due to the identity

\beq
m\frac{d}{dm}\left(\frac{1}{\slashed p-m}\right)=\frac{1}{\slashed p-m}m\frac{1}{\slashed p-m}.
\eeq

\noindent
Then \eq{selfenexpdif} can be written in the form

\beq
\Gamma_m(m) +m\delta Z_2=m\frac{d\Sigma(\slashed p=m)}{dm},
\eeq

\noindent
and \eq{sigmagamrel} turns into ($\delta m^{(2)}=\Sigma(\slashed p=m)$)

\beq \label{massdercon}
\Sigma(m)=m\frac{d\Sigma(\slashed p=m)}{dm}+\gamma_mm\equiv  
m\frac{d(\delta m^{(2)})}{dm}+\gamma_mm.
\eeq

\begin{figure}[h!]
\begin{center}
\includegraphics
[height=1.5cm]
{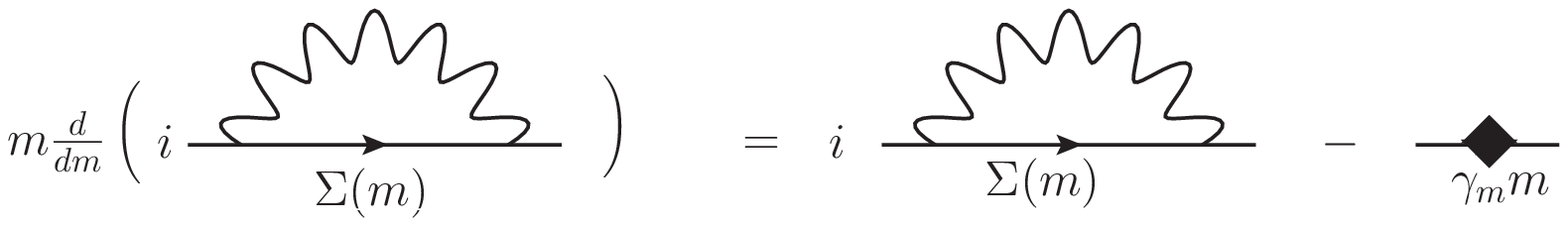}
\end{center}
\caption{Logarithmic mass derivative of $\Sigma(m)$.}
\label{derivm}
\end{figure}

\noindent
We calculate the derivative on the RHS using the explicit expression for $\delta m^{(2)}$ in \eq{mdz2dm} and obtain (see Fig.~\ref{derivm})

\beq \label{relresp}
m\frac{d(\delta m^{(2)})}{dm}
=\delta m^{(2)}-\mu\frac{d\delta m^{(2)}}{d\mu}
=\delta m^{(2)}-\gamma_mm,
\eeq

\noindent
where at the last step we used the definition of the electron mass anomalous dimension. Notice that this relationship holds due to a specific functional form of the mass counterterm $\delta m^{(2)}=mf(m/\mu)$ with some function $f$.

Thus we proved by direct calculation that \eq{massdercon} holds and the expressions in \eq{massnonren} and \eq{sumofford} (and the respective sets of diagrams in Fig.~\ref{mass_ren_stn} and Fig.~\ref{tworelsecc}) coincide.

\section{Conclusions}

We have shown that the standard mass renormalization in Fig.~\ref{mass_ren_stn} and the sum of the diagrams for the matrix element of the EMT trace in Fig.~\ref{tworelsecc} coincide.  This happens due to two important relationships. First, the one-loop  diagram for a scalar vertex is equal to the logarithmic derivative of the self-energy diagram, see \eq{gammdersigm} and Fig.~\ref{deriv}. Second,  the mass renormalization counterterm (self-energy  at $\slashed p=m$) is equal to its own logarithmic derivative plus the product of mass and its anomalous dimension, see \eq{massdercon} and Fig.~\ref{derivm}.

The calculations above are made in the one-loop approximation, but we expect that they, including the functional relationship mentioned after \eq{relresp}, can be generalized to any number of loops. Really, connection between an arbitrary diagram and its logarithmic derivative with respect to the fermion mass, and the connection between such derivative and the fermion mass anomalous dimension do not depend on the number of loops, and these are the only essential steps in the derivation above.

\appendix
\section{Standard one-loop electron mass renormalization}

Some well known results are collected below. We use dimensional regularization and mass-shell renormalization. The QED  Lagrangian in this scheme is

\beq
{\cal L}_0=-\frac{1}{4}F_0^2+\bar\psi_0(i\slashed\partial-m_0)\psi_0
-e_0\bar\psi_0\slashed A_0\psi_0={\cal L}+\delta{\cal L},
\eeq

\noindent
where

\beq
{\cal L}=-\frac{1}{4}F^2+\bar\psi(i\slashed\partial-m)\psi
-\mu^\frac{\epsilon}{2} e\bar\psi\slashed A\psi,
\eeq
\beq
\delta {\cal L}=-\frac{1}{4}\delta Z_3F^2+\bar\psi(i\delta
Z_2\slashed\partial-\delta_m)\psi
-\mu^\frac{\epsilon}{2} e\delta Z_1\bar\psi\slashed A\psi,
\eeq
\beq
{\cal L}+\delta{\cal L}
=-\frac{1}{4}Z_3F^2+iZ_2\bar\psi\slashed\partial\psi-mZ_m\bar\psi\psi
-eZ_1\mu^\frac{\epsilon}{2}\bar\psi\slashed A\psi,
\eeq

\noindent
and

\beq
Z_1=1+\delta Z_1,\quad
Z_2=1+\delta Z_2, \quad
Z_3=1+\delta Z_3, \quad
mZ_m=m(1+\delta Z_m)=m+\delta_m.
\eeq

We define $\delta m= m-m_0=m-mZ_mZ_2^{-1}$. In the one-loop approximation $\delta m^{(2)}\approx -m\delta Z_m+m\delta Z_2=-\delta_m+m\delta Z_2 $.

The renormalized one-loop self-energy $\Sigma_r(p)$ is

\beq \label{renormsigmonshdim}
\begin{split}
\Sigma_r(p)&=
\frac{\alpha}{2\pi}
\int_0^1dx\Biggl\{(2m-x\slashed p)
\left[\frac{2}{\epsilon}-\gamma+\ln(4\pi)+
\ln\frac{\mu^2}{-x(1-x)p^2+x\lambda^2+(1-x)m^2}\right]\\
&-(m-x\slashed p)\Biggr\}-(\slashed p\delta Z_2-\delta_m)_{|\slashed p\to m}\\
&\approx \Sigma(m)+(\slashed p-m)\Sigma'(\slashed p=m)-(\slashed p-m)\delta Z_2+(\delta_m-m\delta Z_2)\\
&=\frac{3\alpha}{4\pi}m
\left[\frac{2}{\epsilon}-\gamma+\ln(4\pi)+\ln\frac{\mu^2}{m^2}
+\frac{4}{3}\right]\\
&-(\slashed p-m)
\frac{\alpha}{4\pi}\left[\frac{2}{\epsilon}-\gamma+\ln(4\pi)
+\ln\frac{\mu^2}{m^2}+2\ln\frac{\lambda^2}{m^2}+{4}\right]\\
&-(\slashed p-m)\delta Z_2+(\delta_m-m\delta Z_2),
\end{split}
\eeq

\noindent
where $\Sigma(p)$ is the dimensionally regularized self-energy, $\mu$ is the auxiliary dimensional regularization mass,  $\lambda$ is the IR photon mass and $\epsilon=4-d$.

\noindent
The one-loop counterterms are

\beq \label{mdz2dm}
\delta m^{(2)}\equiv m\delta Z_2-\delta_m=\Sigma(m)
=\frac{3\alpha}{4\pi}m
\left[\frac{2}{\epsilon}-\gamma+\ln(4\pi)+\ln\frac{\mu^2}{m^2}
+\frac{4}{3}\right],
\eeq
\beq \label{deltaz2}
\delta Z_2=\Sigma'(\slashed p=m)
=-\frac{\alpha}{4\pi}\left[\frac{2}{\epsilon}-\gamma+\ln(4\pi)
+\ln\frac{\mu^2}{m^2}+2\ln\frac{\lambda^2}{m^2}+{4}\right].
\eeq

\acknowledgments

This work was supported by the NSF grant PHY- 2011161.

\end{document}